\documentclass[aps,prl,floatfix,nofootinbib,twocolumn,superscriptaddress]{revtex4}
\usepackage{amsmath, amsfonts, amsthm, amssymb, graphicx}
\usepackage{epstopdf}

\usepackage[all]{xy}
\usepackage{epsfig}
\usepackage{subfigure}

\newcommand{\be}{\begin{equation}}
\newcommand{\ee}{\end{equation}}
\newcommand{\ba}{\begin{eqnarray}}
\newcommand{\ea}{\end{eqnarray}}
\newcommand{\bml}{\begin{mathletters}}
\newcommand{\eml}{\end{mathletters}}

\newcommand{\del}{\ensuremath{\partial}}

\newcommand{\lab}{\label}

\begin{document}

\title{Colliding branes and big crunches.}
\author{John Omotani}
\email[email:]{ppxjto@nottingham.ac.uk}
\author{Paul M. Saffin}
\email[email:]{paul.saffin@nottingham.ac.uk}
\affiliation{School of Physics and Astronomy, University Park, University of Nottingham, Nottingham NG7 2RD, UK}
\author{Jorma Louko}
\email[email:]{jorma.louko@nottingham.ac.uk}
\affiliation{School of Mathematical sciences, University of Nottingham, Nottingham NG7 2RD, UK}

\date{\today}

\begin{abstract}
We examine the global structure of colliding domain walls in AdS spacetime and come to the conclusion that singularities forming from such collisions are of the big-crunch type rather than that of a black brane.
\end{abstract}

\maketitle

\section{Introduction}
\lab{intro}
As one of the two competing theories for hiding extra dimensions, braneworlds have received a lot of attention, starting from the first attempts to get matter to stick on a field theory domain wall \cite{Akama:1982jy,Rubakov:1983bb}, and then to the understanding of how gravity can become localized on a hypersurface \cite{Randall:1999vf}. This naturally led to braneworld cosmological models \cite{Langlois:2002bb} with the possibility that the big-bang may simply correspond to the collision of two brane-worlds \cite{Khoury:2001wf,Bucher:2001it,Gen:2001bx,Langlois:2001uq}. With this in mind there have been a number of numerical studies examining the behaviour both of how matter on the branes react to such collisions \cite{Gibbons:2006ge,Saffin:2007ja,Saffin:2007qa,Takamizu:2004rq}, and to how the spacetime geometry itself deals with the collisions \cite{Takamizu:2006gm,Takamizu:2007ks}. As is to be expected, if the collision of the walls is energetic enough then a singularity will form due to gravitational collapse, this was the main focus of \cite{Takamizu:2007ks} and is what we concern ourselves with here, in particular the global structure of collisions where a curvature singularity forms. The symmetry of the problem is that of parallel branes with three flat, extended spatial directions, so any singularity that forms will have the same symmetry, leading Takamizu {\it et al} \cite{Takamizu:2007ks} to write that the end state of the collision process would be the black hole with these symmetries. In this paper we re-examine the same system, but we claim instead that the end state of a collision process that forms a curvature singularity is in fact a big-crunch, there are no asymptotic regions for any observers to hide in.

\section{field theory domain wall}
\begin{figure}
\centering
\includegraphics[width=7cm]{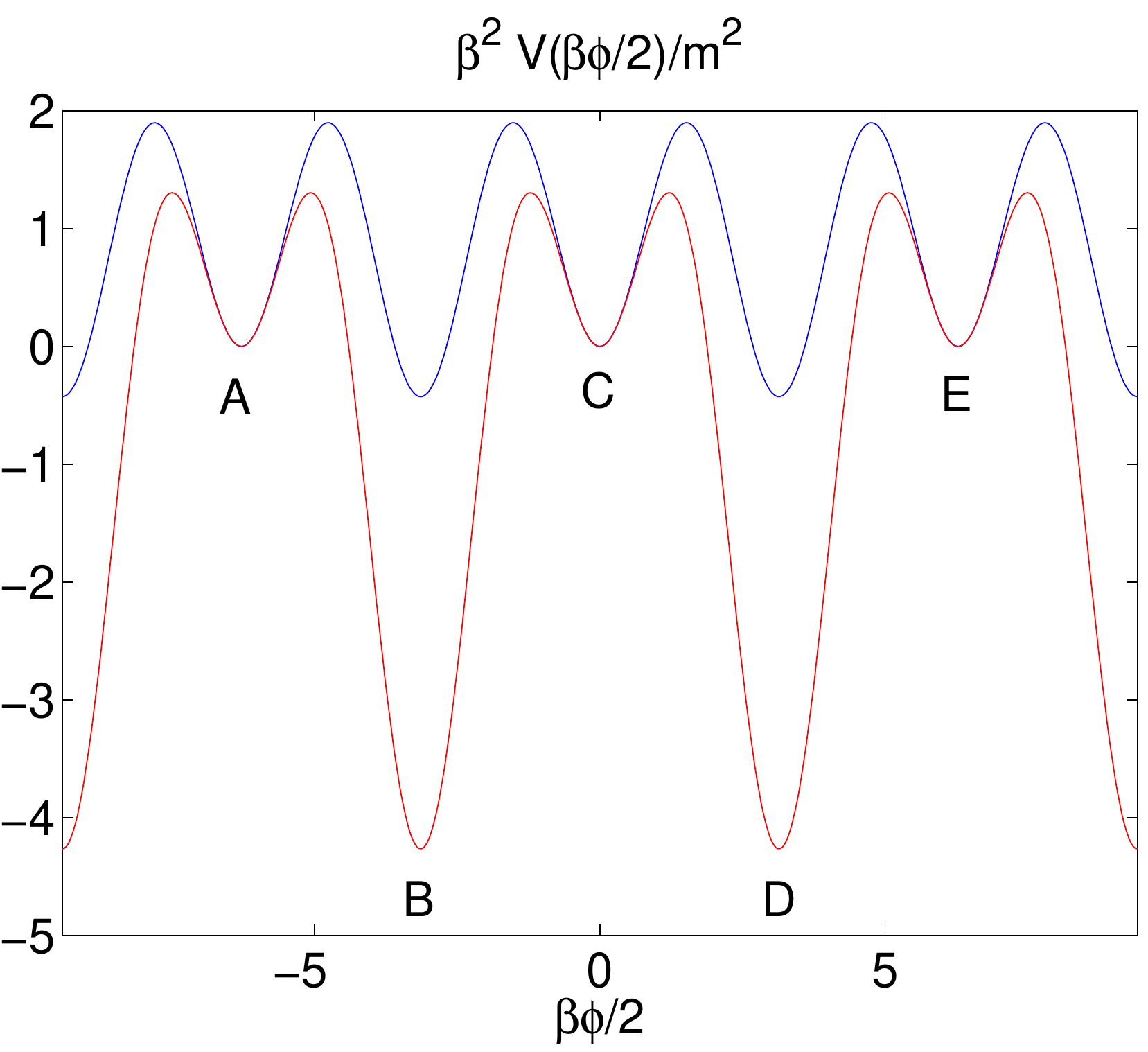}
\caption{\label{fig:potential}The potential for $\beta^2 m_p^3=10$ (lower curve) and $\beta^2 m_p^3=100$ (upper curve). The labels A, B, C, D, E indicate the various minima, with B and D being AdS vacua, and A, C, E being Minkowski.}
\end{figure}

\lab{dwModel}
The model we use to examine the collision of domain walls is that of a single real scalar, canonically coupled to gravity in five dimensions.
\ba
{\cal L}&=&\frac{m_p^3}{2}R-\frac{1}{2}\del_a\phi\del^a\phi-{\cal V}(\phi),
\ea
where we require the potential to have at least two distinct vacua in order for a domain wall solution to exist that can interpolate between them. A rather nice way to achieve this, and one that allows for explicit analytic solutions, is to write the scalar potential in a form inspired by supergravity \cite{Chamblin:1999cj},
\ba
{\cal V}&=&\frac{1}{2}\left[(\del W/\del \phi)^2-\frac{4}{3m_p^3}W^2\right]
\ea
where $W(\phi)$ is termed the superpotential. With this restriction on the form of the potential one finds that a line element and scalar field ansatz of the form
\ba
\label{eq:staticLineElement}
ds^2&=&e^{2U(r)}\eta_{\mu\nu}dx^\mu dx^\nu+dr^2,\qquad\phi=\phi(r),
\ea
leads directly to the following BPS system of equations
\ba
\label{eq:BPSeqns}
\phi'&=&\pm\del W/\del\phi,\\\nonumber
U'&=&\mp\frac{1}{3m_p^3}W.
\ea
where the $\pm$ gives us our kinks or anti-kinks. To be specific we need to make a choice of superpotential, and we pick the sine-Gordon model
\ba
W&=&\mu^4-\frac{4m}{\beta^2}\cos(\beta\phi/2)
\ea
giving a potential of the form
\ba
{\cal V}&=&\left[\frac{2m^2}{\beta^2}-\frac{2\mu^8}{3m_p^3}\right]+\frac{16m\mu^4}{3m_p^3\beta^2}\cos(\beta\phi/2)\\\nonumber
        &~&-\left[\frac{2m^2}{\beta^2}+\frac{32m^2}{3m_p^3\beta^4}\right]\cos^2(\beta\phi/2).
\ea
The three parameters of the superpotential, $m$, $\beta$, $\mu$ have the following effect: firstly, $m$ is the mass of the scalar field if we switch off gravity, i.e. $m_p\rightarrow \infty$, and so $m$ controls the curvature of the potential in the minima; $\beta$ gives the separation of the vacua in field space, with the vacua being located at $\phi_{vac}=n\pi/\beta$; finally, $\mu$ controls the additive constant to the potential, and we shall tune it so that one set of the vacua have vanishing potential, and lead to a Minkowski geometry. To see how to achieve a Minkowski minimum we note that 
\ba
{\cal V}(\beta\phi/2=2\pi n)&=&-\frac{2m^2}{3m_p^3\beta^4}\left(1-\frac{\mu^4\beta^2}{4m}\right)^2,
\ea
and so by choosing $\mu^4=4m/\beta^2$ we have a set of Minkowski minima, leaving us with
\ba\nonumber
&~&{\cal V}=\frac{2m^2}{\beta^2}\left[1-\cos^2(\beta\phi/2)-\frac{16}{3\beta^2m_p^3}[1-\cos(\beta\phi/2)]^2\right]\\
&~&{\cal V}(\beta\phi/2=\pi+2\pi n)=-6m_p^3\left(\frac{8m}{3\beta^2m_p^3}\right)^2=m_p^3\Lambda
\ea
and a form of potential shown in Fig. \ref{fig:potential}, where A, C, E refer to minima tht are Minkowski vacua, and B, D are the AdS$^5$ vacua. In the simulations we perform we shall be using $\beta^2 m_p^3=100$, correspond to the upper curve of Fig. \ref{fig:potential}.

The setup we focus on is the same as Takamizu {\it et al} \cite{Takamizu:2007ks}, where we have two parallel domain walls, with a geometry that asymptotes to AdS$^5$ and contains a Minkowski region sandwiched in between. The field is then taken to interpolate from the B-vacuum, through the C-vacuum, and then on to the D-vacuum. In order to accomplish this we need the profiles of the BC-kink and the CD-kink.

BC kinks are given by the BPS solutions (lower sign of (\ref{eq:BPSeqns}))
\ba\label{eq:kink}
\beta\phi/2&=&2\tan^{-1}\left[\tanh[m(r-r_0)/2]\right]-\pi/2\\\nonumber
U&=&-\frac{4}{3\beta^2m_p^3}\left\{\ln[\cosh[m(r-r_0)]]-\frac{\beta^2\mu^4}{4}(r-r_0)\right\},
\ea
while CD anti-kinks are given by the anti-BPS solutions (upper sign of (\ref{eq:BPSeqns}))
\ba\label{eq:akink}
\beta\phi/2&=&2\tan^{-1}\left[\tanh[m(r-r_0)/2]\right]+\pi/2+\\\nonumber
U&=&-\frac{4}{3\beta^2m_p^3}\left\{\ln[\cosh[m(r-r_0)]]+\frac{\beta^2\mu^4}{4}(r-r_0)\right\}
\ea
What we have just presented are the solutions for the single-kink systems, but we need double kink initial conditions. Although the analytic solution is not available, we are able to add together the BC and the CD kink profiles for $\phi$ and $U$ to provide an excellent approximate solution - so long as the kinks are far enough apart. This is still not quite what we need, as we want to be able to boost the kinks at will, in order to collide them at various speeds; we shall address this when we write down the dynamical equations.

\section{asymptotic structure}
\lab{sec:asStruct}
From the kink solutions (\ref{eq:kink}) (\ref{eq:akink}) we see that the line element (\ref{eq:staticLineElement}) has the asymptotic limit
\ba
ds^2(r\rightarrow+\infty)&\rightarrow& \exp\left[-2\alpha r\right]\eta_{\mu\nu}dx^\mu dx^\nu+dr^2\\
\alpha&=&\frac{8m}{3\beta^2m_p^3}
\ea
and by defining $\alpha Z=\exp\left[\alpha r \right]$ we see that the asymptotic region of the domain wall is given 
\ba
ds^2(Z\rightarrow+\infty)&\rightarrow&\frac{1}{\alpha^2 Z^2}\left[\eta^{\mu\nu}dx^\mu dx^\nu+dZ^2\right]
\ea
which we recognize as a portion of AdS$^5$, in particular, it does not contain the AdS boundary, $Z=0$.

Now let us consider a possible end-state for a singular system generated by the collision of two domain walls. The natural choice isthe AdS$^5$ black-brane given by
\ba
ds^2_{bb}&=&-f(R)dT^2+f^{-1}(R)dR^2+R^2\delta_{ij}dx^idx^j,\\
f(R)&=&-MR^{-2}-\Lambda R^2/6.
\ea
Indeed, a version of Birkhoff's theorem tells us that this is the unique solution with these symmetries \cite{Charmousis:2002rc,Zegers:2005vx} in AdS$^5$ spacetime.
Now note that the asymptotic limit is reached at large $R$, and in order to compare it to the brane case we perform the following co-ordinate transformations, $t=\sqrt{-\Lambda/6}T$, $R=1/(\sqrt{-\Lambda/6}Z)$ and find that the asymptotic region is given by
\ba
ds^2(Z\rightarrow0)&\rightarrow&-\frac{6}{\Lambda Z^2}\left[\eta^{\mu\nu}dx^\mu dx^\nu+dZ^2\right]
\ea
confirming that the asymptotic regions of the domain wall and the asymptotic region of the black brane cover different portions of AdS$^5$. It would therefore be surprising if the the collision of domain walls ended up with a final state that was a black brane.

Before we move on the the dynamical system, we briefly note we use dimensionless variables $\tilde x$, $\tilde\phi$ defined by
\ba
x&=&\tilde x/m,\qquad\phi=\tilde\phi/\beta,
\ea
meaning that we are left with the single physical parameter $\beta^2m_p^3$. From now on we work with the dimensionless variables but drop the tildes. This is analogous to measuring distances and the Planck mass in units of $m$, and measuring $\phi$ in units of $\beta$.
\section{\label{sec:dynamics}the dynamical set-up}
Having found the solutions for isolated, static kinks, we need to know how to get them to move. The simplest way to achieve this is to change to co-ordinates in which the spatial direction defining the wall (the co-dimension one direction), and the time co-ordinate are on an equal footing, in which case there is an explicit SO(1,1) Lorentz symmetry. The metric suited to dynamics is therefore of the form
\ba
ds^2&=&e^{2A(t,z)}(-dt^2+dz^2)+e^{2B(t,z)}\delta_{ij}dx^idx^j,
\ea
with $z$ and $r$ related by $e^{A}dz=\pm dr$ in the static case. The equations of motion using these co-ordinates may be written in a form that highlights the SO(1,1) symmetry as
\ba
\del_{\tilde\mu}\del^{\tilde\mu}\phi+3\del_{\tilde\mu}B\del^{\tilde\mu}\phi&=&e^{2A}\frac{\del {\cal V}}{\del \phi},\\
\del_{\tilde\mu}\del^{\tilde\mu} A-3\del_{\tilde\mu} B\del^{\tilde\mu} B&=&
\frac{1}{m_p^3}\left[-\frac{1}{2}\del_{\tilde\mu}\phi\del^{\tilde\mu}\phi+\frac{1}{3}e^{2A}{\cal V}\right],\\
\del_{\tilde\mu}\del^{\tilde\mu}B+3\del_{\tilde\mu}B\del^{\tilde\mu}B&=&-\frac{2}{3m_p^3}e^{2A},\\
\ea
where the $\hat\mu$ index runs over the $t$, $z$ directions; the constraint equations become
\ba
&~&\del_{\tilde\mu}\del_{\tilde\nu}B
+\del_{\tilde\mu}B\del_{\tilde\nu}B
+\eta_{\tilde\mu\tilde\nu}\del_{\tilde\rho}B\del^{\tilde\rho}B\\\nonumber
&~&-\del_{\tilde\mu}A\del_{\tilde\nu}B-\del_{\tilde\mu}B\del_{\tilde\nu}A
+\eta_{\tilde\mu\tilde\nu}\del_{\tilde\rho}A\del^{\tilde\rho}B\\\nonumber
&=&-\frac{1}{3m_p^3}\left[\del_{\tilde\mu}\phi\del_{\tilde\nu}\phi-\frac{1}{2}\eta_{\tilde\mu\tilde\nu}\del_{\tilde\rho}\phi\del^{\tilde\rho}\phi
                      +\eta_{\tilde\mu\tilde\nu}e^{2A}{\cal V}\right].
\ea
It is now clear that $A(t,z)$, $B(t,z)$ and $\phi(t,z)$ are all Lorentz scalars under this SO(1,1), and so it is easy to boost to different Lorentz frames ${\cal O}$ and ${\cal O}'$ using
\ba\nonumber
t'&=&\gamma(t-vz),\quad z'=\gamma(x-vt),\quad\Psi'(x')=\Psi(x)
\ea
for generic scalars $\Psi$.

Now, because we are interested in studying the global structure of the system, and that system has a singularity, it is actually more convenient to use a double-null co-ordinate system \cite{Takamizu:2007ks}\cite{Burko:1997tb} given by
\ba
u&=&\frac{1}{\sqrt 2}(t-z),\quad v=\frac{1}{\sqrt 2}(t+z)
\ea
because then the null geodesics are simply 45degree lines, and the causal structure is easy to picture. The precise details of the numerical method we use may be found in \cite{Burko:1997tb}, but for another approach see \cite{Martin:2003yh,Frolov:2004rz}.
\section{simulations}
\begin{figure}
\centering
\includegraphics[width=7cm]{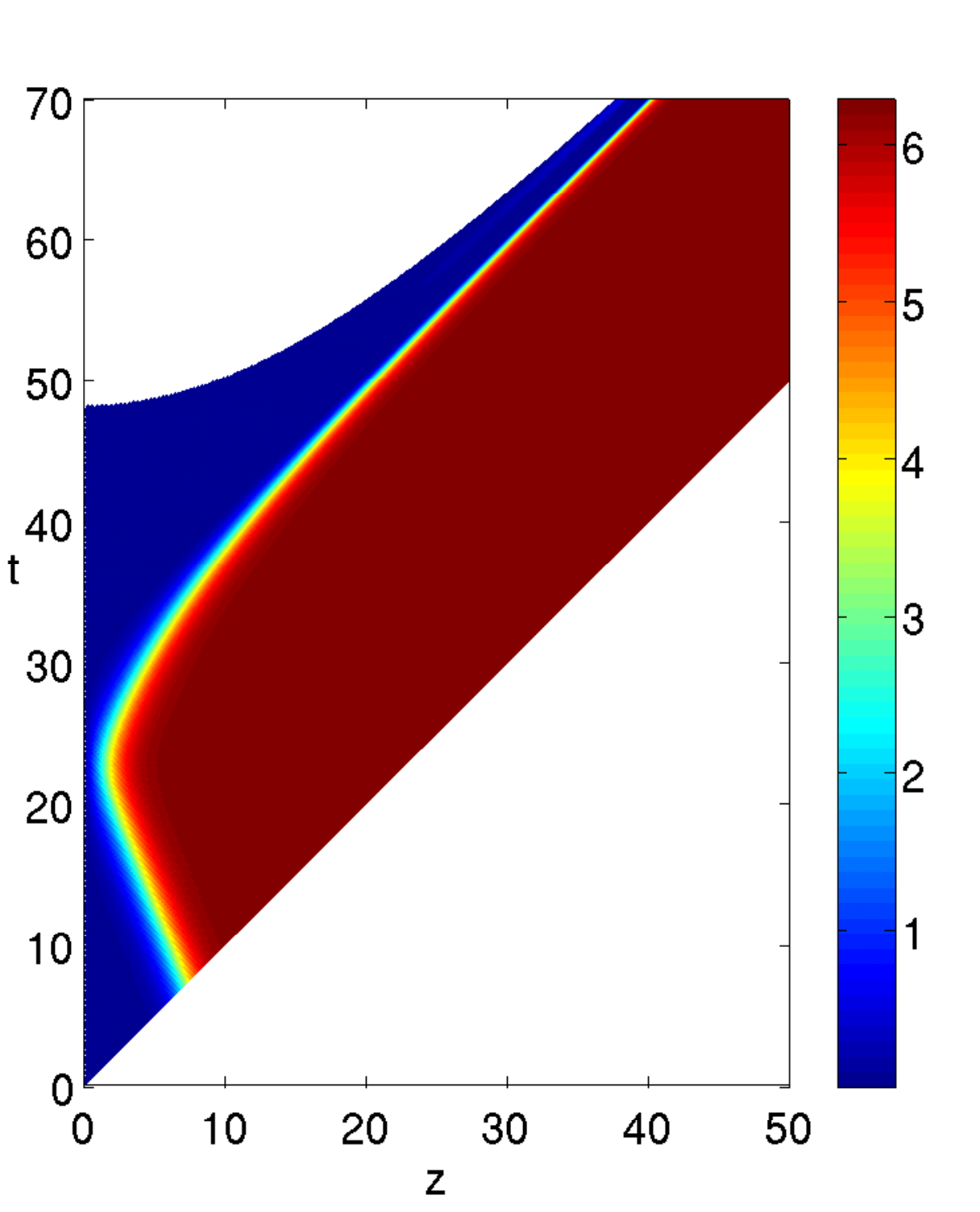}
\caption{\label{fig:phi}Evolution of the scalar field; $\phi=0$ corresponds to the Minkowski minima, and $\phi=2\pi$ to the AdS minima. The upper boundary of the solid-shaded and the white region marks the location of the curvature singularity.}
\end{figure}

\lab{sec:sims}
Having described the system we now move on and give an overview of the collisions that lead to a singularity. To orient ourselves we start with Fig. \ref{fig:phi}, that shows the evolution of the scalar field, where we only simulate the region $z\geq0$, as the $z<0$ region follows by symmetry. Recall that the vacua are $\phi=0,\;2\pi$, and that the $\phi=2\pi$ vacuum corresponds to the AdS$^5$, and the $\phi=0$ vacuum is Minkowski. Fig. \ref{fig:phi} therefore shows two walls coming together, interacting, and then moving apart. 
In the example shown we found that a curvature singularity was forming, and so we cut off the evolution when the curvature became too large. A benefit of the double-null co-ordinates is that one can carry on simulating and map out the region where the curvature gets cut off; in Fig, \ref{fig:phi} this region is given by the upper boundary of the solid-shaded region.
\section{Horizon structure}
\lab{sec:horizon}
Given that a singularity has formed it is natural to ask about the horizon structure of the spacetime, and for this we need to know about the behaviour of null geodesics. It is clear from Fig. \ref{fig:phi} that there is a region inside which timelike geodesics are doomed to end on the singularity, and so we my expect a horizon. However, given the dynamic nature of the system it is actually more convenient to work with objects that have a local definition, namely trapping surfaces. Hayward \cite{Hayward:1993wb} defined trapping surfaces in terms of the expansion of outgoing and ingoing null geodesics, which may be measured without reference to the global properties of the geometry. We start with the co-ordinate vectors $N_+=\del_u$, $N_-=\del_v$ which are ingoing and outgoing respectively (for $z>0$), and introduce their dual one-forms $n_+=-e^{2A}dv$, $n_-=-e^{2A}du$. Normalized outgoing and ingoing null vectors are then defined by $u_\pm=e^{-2A}N_\mp$ such that $n_\pm(u_\pm)=-1$, and an induced three-metric, $h$, is given by $h=g+e^{-2A}n_+\otimes n_-+e^{-2A}n_-\otimes n_+$, where $g$ is the full metric. The expansions are then defined as
\ba
\label{eq:expansions}
\Theta_{\pm}&=&\frac{1}{2}h^{ab}{\cal L}_\pm h_{ab}
\ea
where the Lie derivatives ${\cal L}_\pm$ are taken along $u_\pm$.

\begin{figure}
\centering
\includegraphics[width=7cm]{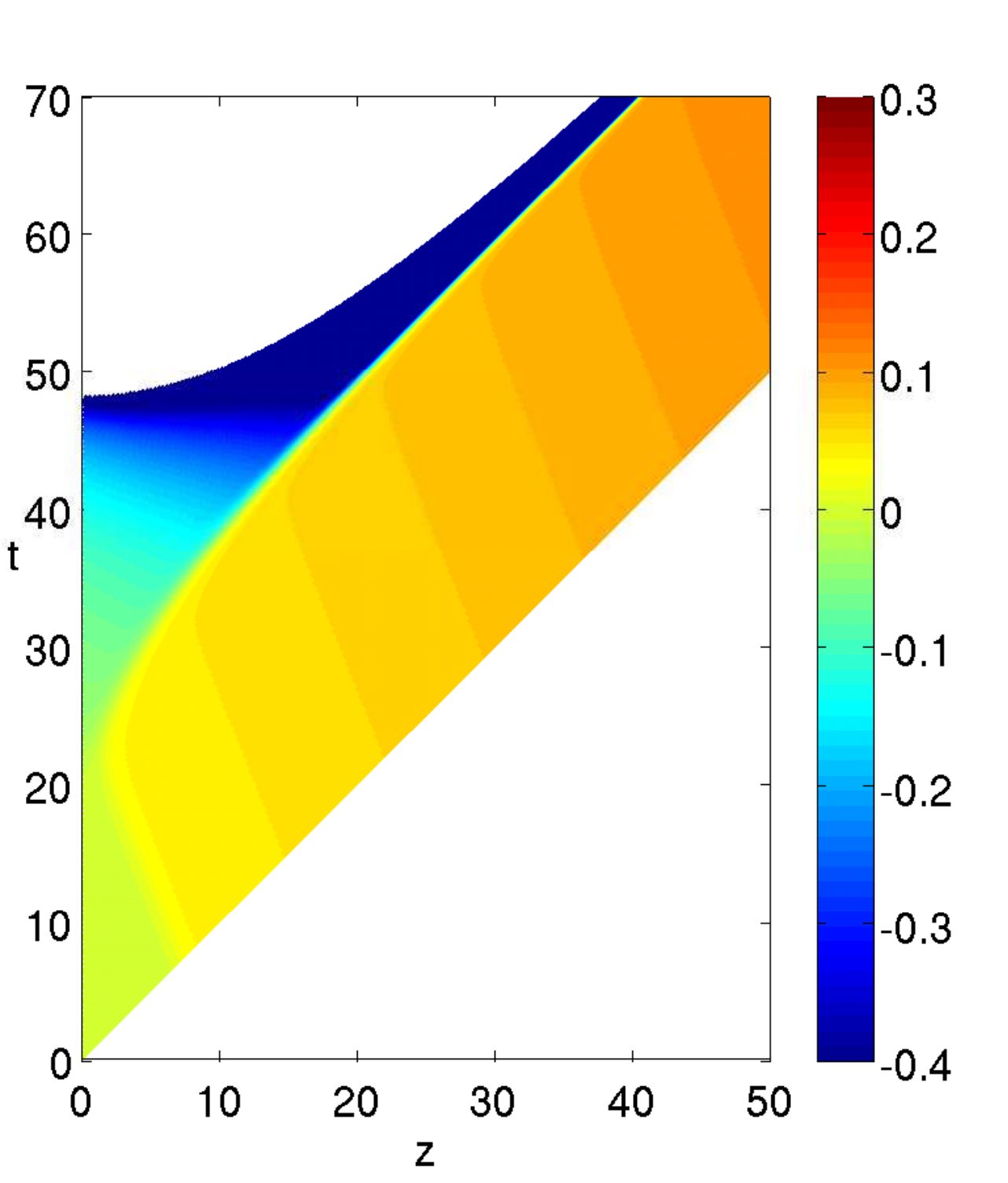}
\caption{\label{fig:thetaMinus}The expansion scalar $\Theta_-$, see (\ref{eq:expansions}).}
\end{figure}

\begin{figure}
\centering
\includegraphics[width=7cm]{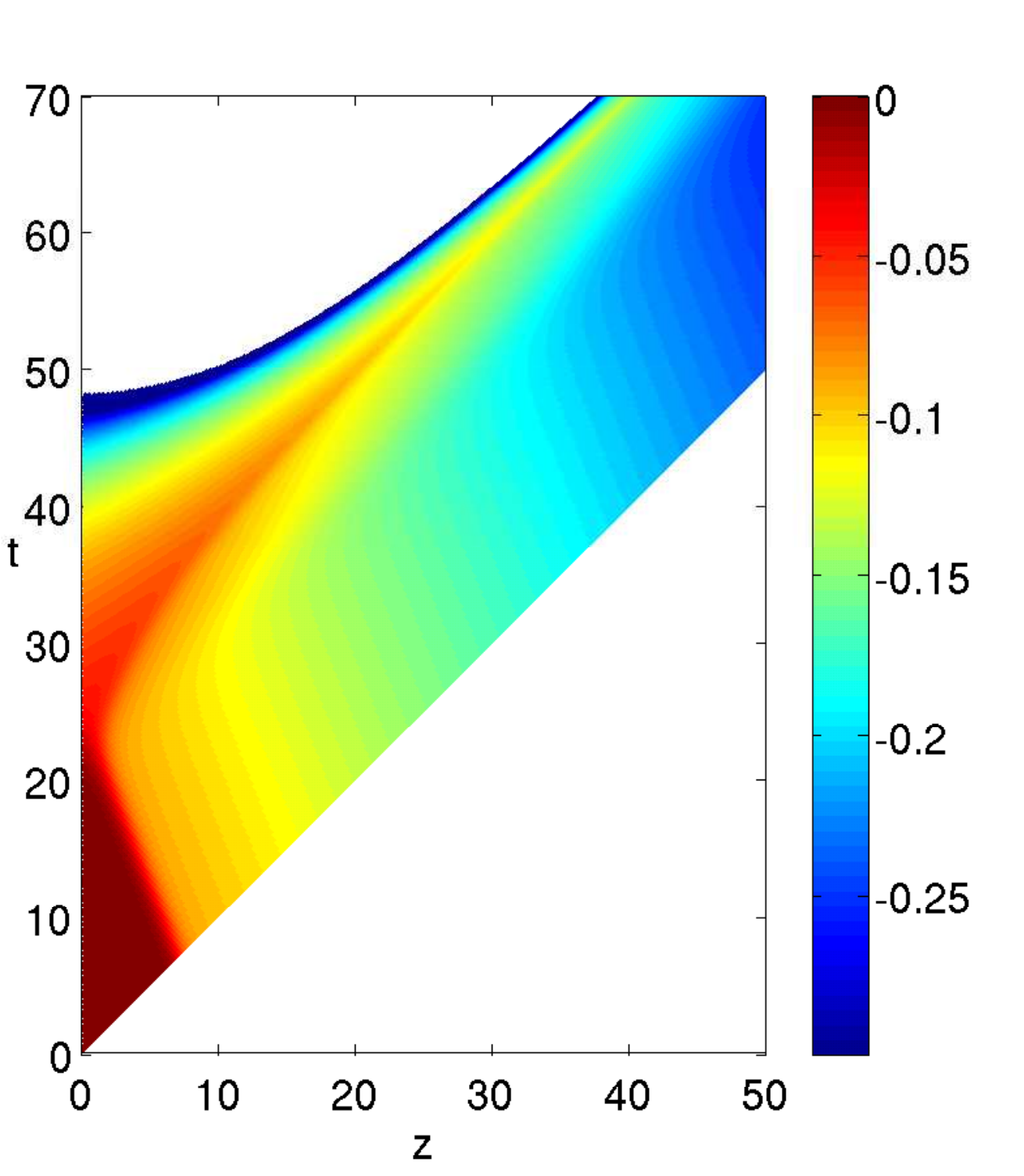}
\caption{\label{fig:thetaPlus}The expansion scalar $\Theta_+$, see (\ref{eq:expansions}).}
\end{figure}

\begin{figure}
\centering
\includegraphics[width=7cm]{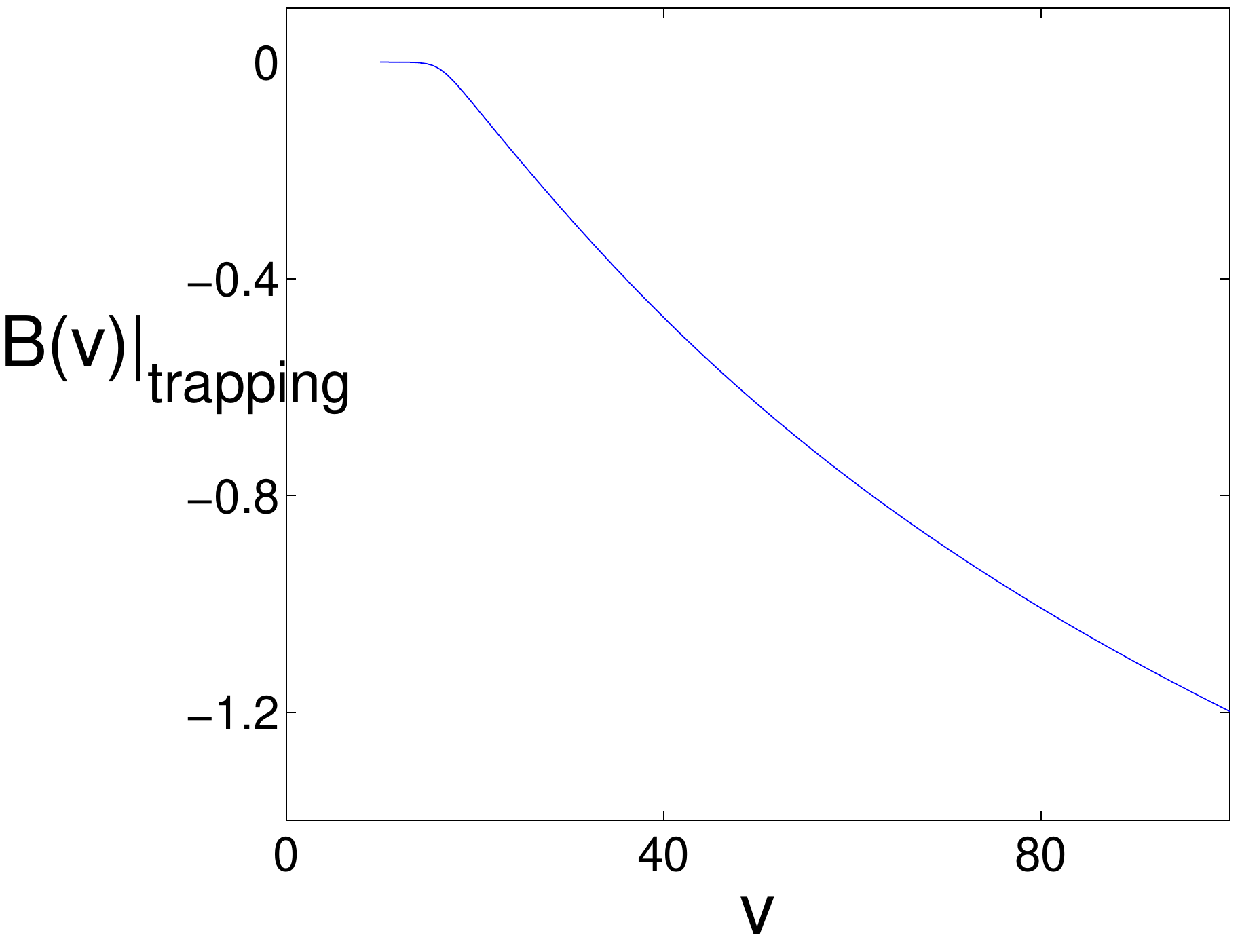}
\caption{\label{fig:B}The value of the metric parameter $B$ as measured on the trapping horizon.}
\end{figure}

A {\it marginal surface} is then a surface where one of the expansions vanishes, say $\Theta_-$. A marginal surface for us is a three-surface, and will be a single point on the $u-v$ plane diagrams such as Fig. \ref{fig:phi}. A trapping horizon is then the four-surface found by sticking together all these marginal surfaces; for us, they correspond to a line on the $u-v$ plane. Having found the trapping surface we can characterize it according to the sign of $\Theta_+$ (the trapping horizon is {\it future} if $\Theta_+<0$ and past if $\Theta_+>0$), and the sign of ${\cal L}_+\Theta_-$ (the trapping horizon is {\it outer} if ${\cal L}_+\Theta_-<0$ and {\it inner} if ${\cal L}_+\Theta_->0$). In Figs. \ref{fig:thetaMinus} and \ref{fig:thetaPlus} we show the expansions, where we clearly see a trapping horizon that separates regions where $\Theta_-$ changes sign, moreover we see from Fig. \ref{fig:thetaPlus} that along this curve $\Theta_+<0$, making it a future trapping horizon, i.e. once you pass this surface your future is determined - you hit the singularity. To see whether the trapping surface is inner or outer we evaluate ${\cal L}_+\Theta_-$, i.e. just see whether $\Theta_-$ increases or decreases along $\del_v$; it increases. That $\Theta_+<0$ and ${\cal L}_+\Theta_->0$ makes the trapping horizon a future inner horizon, which is the same type as one finds in a cosmological big crunch, as opposed to black hole trapping horizons which are future-outer.
Another signature of future inner trapping horizons is that its area is non-increasing, which we confirm by measuring the value of $B$ along the trapping horizon in Fig. \ref{fig:B}.

\begin{figure}
\centering
\includegraphics[width=7cm]{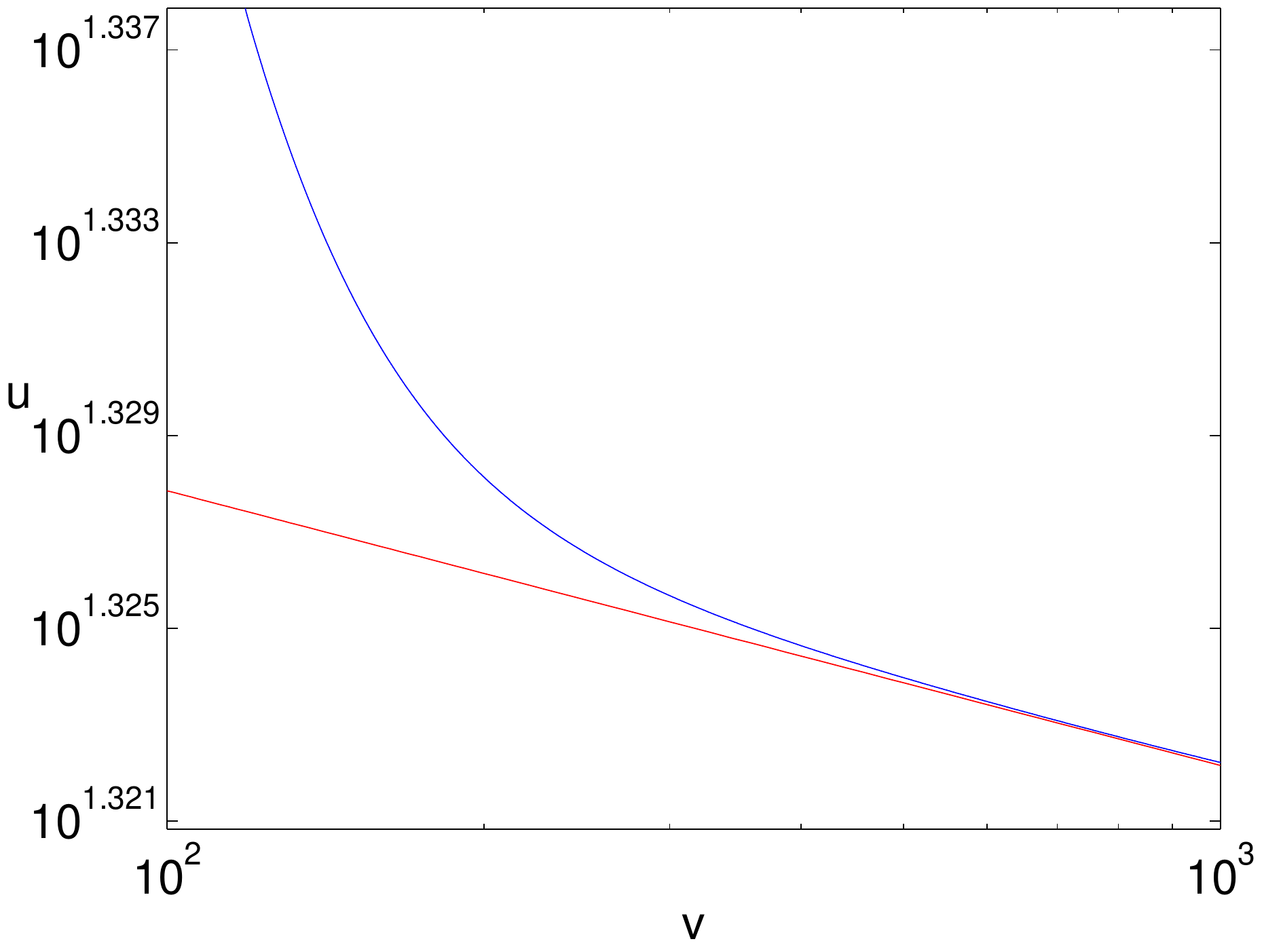}
\caption{\label{fig:ricci}The curved line is the location of Ricci level set $R=-5$, and the straight line is a fit of the form $1/v^{5.7\times10^{-3}}$}
\end{figure}

To really check the claim that what we have is a big crunch, with no asymptotic region we should examine how the singularity behaves in the large-$v$ region. This is clearly a challenging task numerically, but what we can show is the location of level surfaces of the Ricci scalar, with the aim of showing that it cuts across any putative asymptotic region. In Fig. \ref{fig:ricci} we give the location of some level set (in this example it is $R=-5$) and we see that it is consistent with the line hitting $u=0$, albeit rather slowly in these co-ordinates. If this behaviour is repeated for larger values of the Ricci scalar, in particular the singular value, we see that the spacelike singularity cuts off the asymptotic region, ending the spacetime in a big crunch.

\section{conclusions}
\lab{sec:concs}
We have re-considered the analysis of Takamizu {\it et al} \cite{Takamizu:2007ks} with the aim of understanding the global structure of domain wall collisions that form a curvature singularity. By examing the asymptotic regions of domain walls and black branes, and by measuring the behaviour of null rays in the dynamical geometry we conclude that the horizon structure is more consistent with that of a big-crunch, rather than black-brane end-state. Moreover, by following the location of level set of the Ricci scalar we find tentative agreement of this picture. The rather slow fall-off of makes it difficult to track the level sets to sufficient distance in $v$ using these co-ordinates, before numerical error becomes a problem. This can be compared to a prediction of \cite{Chamblin:1999cj} which is that the AdS$^5$ Cauchy horizon, generically gets replaced by a pp singularity when the AdS region is perturbed. Here we claim that in the cases where the collisions form a curvature singularity, then that curvature singularity closes off the geometry and no pp singularity would form. However, in less  violent cases where no curvature singularity is observed, the Cauchy horizon could still be expected to form a pp singularity.

\begin{acknowledgments} 
The authors would like to acknowledge support from STFC.
\end{acknowledgments}


\end{document}